\documentclass[prl,twocolumn,superscriptaddress]{revtex4-2}
\usepackage{graphicx}
\usepackage{dcolumn}
\usepackage{amssymb} 
\usepackage{bm}
\usepackage{pifont}
\usepackage{epsfig}
\usepackage{psfrag}
\usepackage[usenames]{color}

\usepackage[hidelinks]{hyperref}
\hypersetup{
  colorlinks   = true, 
  urlcolor     = blue, 
  linkcolor    = blue, 
  citecolor   = blue 
}

\begin{document}

\title{Anisotropic Cage Evolution in Quasi-two-dimensional Colloidal Fluids}

\author{Noman Hanif Barbhuiya}
\affiliation{Department of Physics, Indian Institute of Technology Gandhinagar, Palaj, Gandhinagar, 382055, Gujarat, India}

\author{Chandan K. Mishra}
\email{chandan.mishra@iitgn.ac.in}
\affiliation{Department of Physics, Indian Institute of Technology Gandhinagar, Palaj, Gandhinagar, 382055, Gujarat, India}

\date{\today}
\draft
\begin{abstract}

We experimentally explore the morphological evolution of cages in quasi-two-dimensional suspensions of colloidal fluids, uncovering a complex dynamic restructuring in the fluid. Although cages display isotropic evolution in the laboratory frame, we observe a striking anisotropy when analyzed in the displacement frame of the caged particles. Moreover, our findings reveal that particles in specific but distinct regions of the cage predominantly contribute to either its persistence or relaxation. Thus, our study provides a coarse-grained microscopic picture of the structural relaxation of these fluids through cage evolution, which has broader implications for the flow and phase behavior of complex fluids in confined geometry.
\end{abstract}
\maketitle
\renewcommand{\thefootnote}

Cage formation, whether transient or long-lived, is a fundamental feature of condensed matter systems such as (supercooled) liquids \& glasses \cite{reis2007caging, gallo1996slow, sahu2024structural, li2020anatomy, weeks2002properties} and crystals \cite{rabani1999direct, quinn2002particle}. Microscopic insights into the structural configuration and dynamical evolution of cages have offered crucial understandings of diverse phenomena, such as the rheological \cite{mayer2008asymmetric} and mechanical properties \cite{li2020anatomy} of the materials as well as about dynamical phases of a system \cite{debets2021cage, zhang2024anisotropic}. For example, the distinct nature of glass transition and prediction of crystal melting in two dimensions (2D) compared to three dimensions (3D), becomes dimension-agnostic when the dynamics of particles relative to their respective cages, rather than their self-dynamics, are considered \cite{illing2017mermin, vivek2017long, zheng1998lindemann, Shiba_2018, flenner2015fundamental}. This is because long-wavelength Mermin-Wagner fluctuations, which influence the dynamics in 2D systems, are believed to be accounted for in the cage-relative perspective \cite{illing2017mermin, vivek2017long, zheng1998lindemann, Shiba_2018}. 

Interestingly, recent studies suggest that long-wavelength Mermin-Wagner fluctuations are also present in 2D colloidal liquids and lead to the violation of the ubiquitous Stokes-Einstein (SE) relation; SE relation connects the microscopic diffusivity of the tracers and the bulk viscosity of the liquid \cite{li2019long}. Once again, the cage-relative dynamics of the particles have been shown to restore the usual behavior of the Stokes-Einstein relation for 2D liquids \cite{li2019long}. However, despite the extensive exploration of cage relative dynamics, the evolution of the cages themselves in 2D colloidal fluids $-$ their transient nature leading to possible changes in their shape over time, which consequently dictates the structural relaxation of the fluid$-$ has never been explored. It is particularly intriguing as long-wavelength fluctuations have been alluded to stem from hydrodynamic interactions \cite{li2019long, shiba2019local}, that are direction-dependent in the body frame of particle-pair for quasi-two-dimensional (q2D) colloidal fluids \cite{barbhuiya2023direction}. Thus, the coarse-grained depiction of fluid dynamics can elucidate the structural relaxation of fluids in confined geometries by examining the dynamic restructuring of cages under emergent hydrodynamic interactions.

\begin{figure}[!h]
\includegraphics[width=0.45\textwidth]{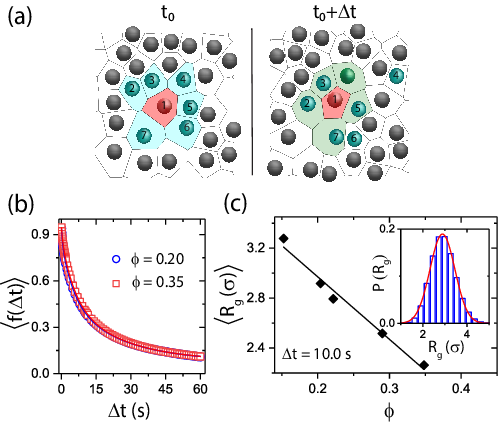}
\caption{Cage characterization in quasi-two-dimensional colloidal fluids. (a) Rendering of the Voronoi cells (polygons) for a representative subset of particles (spheres) in the experimental field of view at initial, $t_0$ (left), and later time, $t_0 + \Delta t$ (right) at $\phi = 0.35$, with $\Delta t = 2.5$ s. The representative reference caged-particle's (red sphere) Voronoi cell with its adjoining cells at $t_0$ (cyan polygons) and $t_0 + \Delta t$ (green polygons), determines the identity of the particles forming the cage (shown as numerals on the spheres). (b) Average fraction of common cage particles, $\langle f(\Delta t) \rangle$, versus lag time, $\Delta t$, for area fractions, $\phi = 0.20$ (blue circles) and $\phi = 0.35$ (red squares). (c) The average radius of gyration of the cages, $\langle R_g \rangle$, versus $\phi$ for $\Delta t = \tau_{\text{cage}}$ with the solid line as a guide to the eye. Inset shows the probability distribution of $R_g$, $P(R_g)$, for $\phi = 0.20$ and $\Delta t = \tau_{\text{cage}}$, with the red curve being a Gaussian fit to the data.}
\label{Figure1}
\end{figure}

\begin{figure}[htbp]
\includegraphics[width=0.45\textwidth]{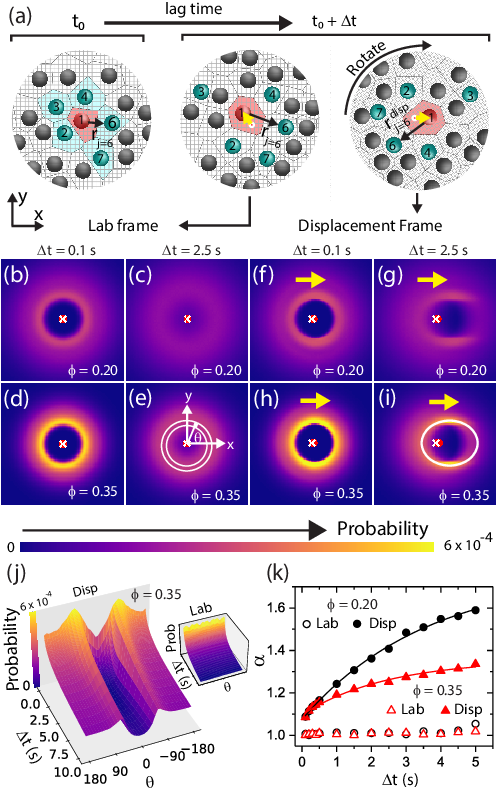}
\caption{Cage evolution in quasi-two-dimensional colloidal fluids. (a) Rendering from experiments illustrating the laboratory and displacement frames of reference. A portion of the field of view at times $t_0$ (left) and $t_0 + \Delta t$ (middle), a representative particle (red sphere) and its cage particles (cyan spheres) are determined from Voronoi cells at $t_0$. The position vectors of the cage particles with respect to the reference caged-particle ($\{\textbf{r}_j\}$) and the displacement of the caged-particle ($\Delta \textbf{r}_i(\Delta t)$) are shown in black and yellow arrows, respectively. The right panel demonstrates the transformation of particle coordinates to the displacement frame, where the displacement of the reference-caged particle (red) is aligned with the x-axis. Ensemble-averaged colormap of $P(\textbf{r}_j(t_0+\Delta t) \vert c_j \in C(i,t_0))$, in the laboratory frame, (b)-(e) , and $P(\textbf{r}_j^{disp}(t_0+\Delta t) \vert c_j \in C(i,t_0))$ in displacement frame, (f)-(i), for two distinct $\phi$ and $\Delta t$. The white cross and red solid circle in (b)-(i) represent the mean of the probability distribution at $t_0$ and $t_0 + \Delta t$, respectively. The yellow arrows in (f)-(i) show the displacement direction of the reference caged-particle. (j) Total probability within an annular region with $1.0 \sigma \leq r \leq 1.25 \sigma$ and centered on the mean of the probability distribution, as a function of angle $\theta$ (as illustrated in (e)) and $\Delta t$ for $\phi = 0.35$, in the displacement frame and laboratory frame (inset). (k) The aspect ratio, $\alpha$, of the fitted ellipse along the maxima of the probability distribution as a function of $\Delta t$ for two distinct $\phi$ in the laboratory (open symbols) and displacement frames (solid symbols).}
\label{Figure2}
\end{figure}

\begin{figure}[htbp]
\includegraphics[width=0.45\textwidth]{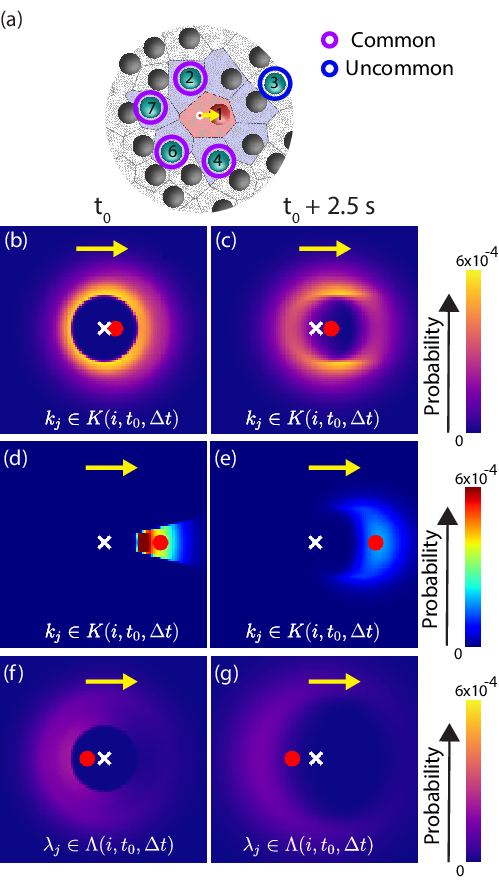}
\caption{Distinct anisotropic spatial probability distribution of the cage particles leading to cage persistence and relaxation. (a) Rendering from the experiments of a portion of the field of view, transformed to its respective displacement frame, illustrating the classification of cage particles (cyan spheres) at time $t_0$ into common, $K(i, t_0, \Delta t)$, and uncommon, $\Lambda(i, t_0, \Delta t)$, subsets based on their presence or absence in the cage defined for the same reference particle (red sphere) a later time, $t_0 + \Delta t$. The common and uncommon cage particles are circled in purple and blue, respectively. Colormap of (b) $P(\textbf{r}_j^{disp}(t_0) \vert k_j \in K(i, t_0, \Delta t))$ and (c) $P(\textbf{r}_j^{disp}(t_0 + \Delta t) \vert k_j \in K(i, t_0, \Delta t))$, at $t_0$ and $t_0 + \Delta t$, respectively, for $\phi = 0.35$ and $\Delta t = 2.5$ s. (d) and (e) are the same as (a) and (b), respectively, but considering only the subset of particles within an angular region of $\pm 15^{\circ}$ with respect to the direction of displacement of the reference-caged particle. (f) and (g) depict the same as (b) and (c) but for the uncommon cage particles, $\Lambda(i, t_0, \Delta t)$. The white cross and red solid circle in (b)-(g) represent the mean of the probability distribution at $t_0$ and $t_0 + \Delta t$, respectively. The yellow arrows in (b)-(g) show the displacement direction of the reference caged-particle.}
\label{Figure3}
\end{figure}

In this Letter, we employ video microscopy to experimentally investigate the spatiotemporal evolution of particles forming the cages in quasi-two-dimensional (q2D) colloidal fluids, both in the laboratory and in the displacement frame of the reference caged-colloid. As expected, the fraction of particles that remain part of the reference caged-particle, defined at the initial time, $t_0$, decreases exponentially over time, facilitating the structural relaxation of the fluid. Notably, the observed size of cages over the cage-relaxation timescales suggests that their morphological evolution is strongly influenced by near-field hydrodynamic interactions in q2D confinement [Figure \ref{Figure1}] \cite{barbhuiya2023direction}. As a result, while the mean shape of the cages appears to evolve isotropically in the laboratory frame of reference, their evolution is anisotropic and asymmetric in the displacement frame of the reference caged-particle [Figure \ref{Figure2}]. Finally, consistent with hydrodynamic motional modes, we identify distinct hotspots within the initial cage structure where particles either contribute to cage persistence or facilitate cage relaxation, thereby unravelling the crucial role of cages in the structural relaxation of the fluid [Figure \ref{Figure3}].

Aqueous suspension of charge-stabilized polystyrene colloidal particles, diameter: $\sigma = 1.04$ $\mu$m, polydispersity $\sim 3\%$, were loaded in a wedge-shaped cell, which was then allowed to stand vertically for particles to sediment under gravity and form a monolayer in the quasi-two-dimensional region of the cell \cite{barbhuiya2023direction}. Once a desired area fraction was achieved, the cell was equilibrated for several hours under the microscope. After equilibration, video microscopy was performed for 20 minutes using 100x oil objective (1.4 Numerical aperture) at 10 frames per second (fps) using the Hamamatsu ORCA-Flash4.0 camera. We have captured data in the liquid regime at five area fractions, $\phi$, in the range, $0.15 \leq \phi \leq 0.35$, in the same region of the cell. The trajectory of each particle was determined using standard image processing and tracking algorithms \cite{crocker1996methods}. The dynamic spatial resolution in our experiments was 20 nm. 

The identity of cage particles, $C(i,t)$, for $i^{th}-$particle at a time, $t$, is determined using Voronoi tessellation, which partitions the 2D space into neighborhoods (cells) around each particle such that every point within a cell is closer to the particle than to any other particle [Fig. \ref{Figure1}(a)] \cite{burrough2015principles}. The order of $C(i,t)$, $n(C(i,t))$, represents the number of particles forming the cage of the $i^{th}-$particle that is also equal to the number of vertices in its Voronoi cell. With time, the identity of particles contained in $C(i,t)$ changes, leading to the relaxation of the cage and, consequently, the structural relaxation of the fluid. 

The relaxation (or persistence) time of the cage is measured by calculating the ensemble-averaged fraction of common cage particles, $\langle f(\Delta t) \rangle$, with lag time, $\Delta t$, $\langle f(\Delta t) \rangle = {1\over N} {\langle \sum_{i}^N \frac{n(C(i, t_0) \cap C(i, t_0 + \Delta t))}{n(C(i, t_0))} \rangle}_{t_0}$. Here, ${\langle \cdot \rangle}_{t_0}$ represents initial time averaging over $t_0$, corresponding to a given $\Delta t$, and $N$ is the total number of particles. $\langle f(\Delta t) \rangle$ exhibits a stretched-exponential decay and yields the persistence (or relaxation) time of the cage, $\tau_{cage}$ [Fig. \ref{Figure1}(b)]. For the range of $\phi$s studied, $\tau_{cage} \sim 10$ s, and surprisingly, does not seem to change with $\phi$. This suggests that coarse-grained structural relaxation of the fluid over the lengthscales of the cages, typically associated with their bulk relaxation, maybe density-independent. 

Next, to probe the spatial extent or lengthscales associated with cage relaxation, we have measured the radius of gyration, $R_g$, of the cages at $\Delta t = \tau_{cage}$. $R_g(i, \Delta t)$, measured from the positions of particles at $t_0 + \Delta t$, which forms the cage of $i^{th}-$particle at $t_0$, and is defined as, $\displaystyle R_g(i, \Delta t) = \sqrt{{1\over n(C(i, t_0))} \sum_{c_j \in C(i, t_0)} (\textbf{r}_j(t_0 + \Delta t) - {\langle \textbf{r}_j(t_0 + \Delta t) \rangle}_j})^2$. Here, $\textbf{r}_j(t_0 + \Delta t)$ is the positions of the $j^{th}-$particle belonging to the set of particles that form the cage of $i^{th}-$particle at $t_0$. Note, throughout the Letter, the sets and their elements are represented using upper and lowercase alphabets, respectively. The average radius of gyration of the cages, $\langle R_g \rangle$, expectedly decreases with increasing $\phi$ [Fig. \ref{Figure1}(c)]. More importantly, the average cage size over the $\tau_{cage}$ turns out to be $\sim 2-3 \sigma$ [Fig. \ref{Figure1}(c)], implying that the temporal evolution of cages should be strongly governed by the nature of near-field hydrodynamic interactions in q2D  spatial confinement, which is direction-dependent (anisotropic) in the body frame of colloid-pairs \cite{barbhuiya2023direction}.  

As the fluid relaxes, particles of the cage, as defined at $t_0$, will diffuse and may lead to a change in the shape of the cage. Thus, to capture the microscopic evolution in the shape of the cages with $\Delta t$, we plot the probability distribution, $P({\textbf{r}_j(t_0+\Delta t)} \vert {c_j \in C(i,t_0)})$, of the positions of the particles of the cage, defined at $t_0$, over $\Delta t$ [Fig. \ref{Figure2}(a)-(e)]. Here, the argument on the right of the ``$\vert$'' represents the condition applied for the quantity being measured on the left. Note that the probabilities are calculated from the ensembles of particles (and their cages) and, for a given $\Delta t$, averaged over $t_0$. Particles forming the cage spread radially, both outward and inward, from their initial position, with $P(\textbf{r}_j(t_0+\Delta t) \vert c_j \in C(i,t_0))$ being symmetric about the center of mass of the cages at $t_0$. The diffused annular ring signifies the caging of the particle by its neighbors in colloidal fluids [Fig. \ref{Figure2}(b)-(e)], and as expected, the annular ring spreads with $\Delta t$. For a given $\Delta t$, with an increase in $\phi$, as caging effects strengthen, the annular ring becomes more localized [Fig. \ref{Figure2}(b)-(i) \& Supplemental Video 1]. 

$P(\textbf{r}_j(t_0+\Delta t) \vert c_j \in C(i,t_0))$ is found to be symmetric. However, note that the laboratory frame of measurements of the positions of particles forming the cage for the analysis of $P(\textbf{r}_j(t_0+\Delta t) \vert c_j \in C(i,t_0))$ will be unable to capture any possible influence of the unique nature of hydrodynamic interactions in q2D colloidal fluids on the dynamic restructuring of the cages. Hence, to discern the effects of hydrodynamics on the evolution of the shape of the cages, we calculate $P(\textbf{r}_j^{disp}(t_0+\Delta t) \vert c_j \in C(i,t_0))$ in the displacement frame of the $i^{th}-$reference caged-particle. Here, we rotate the positions of all the particles of the cages such that the displacement of reference-caged particle aligns along the $x-$axis [Fig. \ref{Figure2}(a)]. In essence, we transform $\textbf{\textbf{r}}_i$ in the lab frame to $\textbf{r}_i^{disp}$ in displacement frame by $\textbf{r}_i^{disp} = \textbf{R}(-\psi) \textbf{r}_i$, where $\psi$ is angle that the displacement of the reference caged-particle subtends with $x-$axis and $\textbf{R}(-\psi)$ is the rotation matrix with elements $R_{lm}(\psi) =\cos(\psi)\delta_{lm} + \sin(\psi)\delta_{lm}$.       

Interestingly, unlike $P(\textbf{r}_j(t_0+\Delta t) \vert c_j \in C(i,t_0))$, $P(\textbf{r}_j^{disp}(t_0+\Delta t) \vert c_j \in C(i,t_0))$ deviates from the symmetric annular shape and is no longer isotropic with probability densities being different in distinct regions [Fig. \ref{Figure2}(f)-(i) \& Supplemental Video 1]. To quantify this localization of the probabilities, we calculate the net probability within an annular region with an inner and outer radius of $1\sigma$ and $1.25 \sigma$, and centred on the mean of the probability distribution, as a function of angle $\theta$ (for example, see Fig. \ref{Figure2}(e)). For a fixed $\phi$, while the net probability is constant for all $\theta$ for $P(\textbf{r}_j(t_0+\Delta t) \vert c_j \in C(i,t_0))$, however, it displays two maxima at $\theta \sim \pm 90$ for $P(\textbf{r}_j^{disp}(t_0+\Delta t) \vert c_j \in C(i,t_0))$ [Fig. \ref{Figure2}(j)]. It indicates an accumulation of particles forming the cage in regions perpendicular to the motion of the caged-particle, which is in concordance with the dipolar nature of hydrodynamics in q2D colloidal fluid.

To further quantify the anisotropy in $P(\textbf{r}_j^{disp}(t_0+\Delta t) \vert c_j \in C(i,t_0))$, we fit an ellipse along the maxima of the probability distribution \cite{supplemental}. We define the aspect ratio of the fitted ellipsoid, $\alpha$, the ratio of the major to the minor axis, as the anisotropy parameter associated with the probability distribution. In the laboratory frame, irrespective of the $\phi$ and $\Delta t$, $\alpha$ remains closer to unity, suggesting the isotropic spread of the particles of the cages around the reference caged-particle [Fig. \ref{Figure2}(k))]. However, intriguingly, in the displacement frame, as already visually inferred from the colormaps in Figure \ref{Figure2}(f)-(i), the shape anisotropy parameter $\alpha$ increases with $\Delta t$. The angle-dependent localization of $P(\textbf{r}_j^{disp}(t_0+\Delta t) \vert c_j \in C(i,t_0))$ and its time-dependent elongation, thus suggests a strong role of hydrodynamics in governing the evolution of cages, potentially dictating the mechanisms of cage breaking (and persistence).

Finally, having established that the shape of the cage is anisotropic in the displacement frame of the reference caged-particle, we now delve into how the particles of the cage rearrange, which either leads to the persistence of the cage or its breaking and local structural relaxation of the fluid \cite{chen2023visualizing, cao2017release}. In other words, are there any specific regions in the cage that predominantly help with cage relaxation and persistence? Note, for a reference caged-particle, say $i^{th}-$particle, and its cage particles $C(i, t_0)$, defined at time $t_0$, a subset of these particles, $K (i, t_0, \Delta t)$, will remain part of the cage $C(i, t_0 + \Delta t)$ at later time $t_0 + \Delta t$. This common subset, $K (i, t_0, \Delta t) = C(i, t_0) \cap C(i, t_0 + \Delta t)$, contributes to cage persistence, while particles that were part of $C(i, t_0)$ but are not included in $C(i, t_0 + \Delta t)$, denoted as $\Lambda(i, t_0, \Delta t) = C(i, t_0) \setminus C(i, t_0 + \Delta t)$, contribute to cage breaking [Fig. \ref{Figure3}(a)]. Here, ``$\setminus$" denotes the set difference. 

Thus, to investigate if particles in certain spatial regions of the cage are more likely to contribute to cage persistence, we analyze $P(\textbf{r}_j^{disp}(t_0) \vert k_j \in K(i,t_0, \Delta t))$ and $P(\textbf{r}_j^{disp}(t_0+\Delta t) \vert k_j \in K(i,t_0, \Delta t))$. We find that at $t_0$, particles that are located in the direction of the displacement of the reference caged-particle have a higher likelihood of remaining part of the cage of the same particle at $t_0 + \Delta t$ [Fig. \ref{Figure3}(b) \& Supplemental Video 2]. At $t_0 + \Delta t$, however, the common particles of the cage at $t_0$ and $t_0 + \Delta t$ are more likely to be found in the transverse direction of the reference caged-particle [Fig. \ref{Figure3}(c)  \& Supplemental Video 2]. 

To further corroborate these arguments, we have only considered a subset of $K(i, t_0, \Delta t)$ such that they lie within an angular region of $ \pm 15^{\circ}$ with respect to the direction of displacement of the reference caged-particle. Figure \ref{Figure3}(d) and (e) clearly show that common cage particles rearrange in a distinctive ``mushroom cloud" pattern, rearranging their positions vertically and moving opposite to the direction of the reference caged-particle's displacement. Conversely, the uncommon particles of the cage, initially distributed opposite to the displacement direction at $t_0$, become more diffused at $t_0 + \Delta t$ [Fig. \ref{Figure3}(f)-(g)  \& Supplemental Video 2]. This indicates that cage breaking and structural relaxation are predominantly facilitated by particles of the cage that are in the regions opposite to the reference caged-particle's motion. 

In summary, our experiments represent a first-of-its-kind unravelling of the coarse-grained microscopic dynamics of rigidly confined colloidal fluids, focusing on the dynamic restructuring of particle cages. We demonstrate that, in the laboratory frame of reference, cages evolve isotropically and symmetrically. However, in the displacement frame associated with the motion of the reference caged-particle, the evolution dynamics of the cages are markedly different, becoming anisotropic and asymmetric. Particles that initially form the cage and are positioned ahead of the reference particle reorganize and diffuse transversely to the displacement direction of the caged-particle, while still remaining a part of the cage at a later time, thereby contributing to the persistence of the cage. Conversely, particles behind the reference caged-particle tend to diffuse further away, primarily leading to cage breaking and facilitating the structural relaxation of the fluid.

Intriguingly, our findings show that the cage relaxation timescales, determined solely by the persistence of the fraction of cage particles over time, remains independent of the packing area fractions examined in this study. It would be valuable to investigate whether this trend persists in colloidal suspensions at even higher densities, specifically for $\phi > 0.35$, or if a potential dynamic phase transition occurs within the liquid regime. Additionally, considering contrasting strengths, which is $\phi-$dependent, and the phase difference between different hydrodynamic motional modes in q2D \cite{barbhuiya2023direction}, this suggests the possibility of different mechanisms of structural relaxation at different $\phi$ for rigidly confined colloidal fluids. While our study focused on cage dynamics in q2D rigid confinement, it would be worthwhile to explore the coarse-grained dynamics at the lengthscales of cages in suspensions of particles at interfaces \cite{illing2017mermin,van2023tuning} and those confined on curved surfaces \cite{bausch2003grain, singh2022observation}. Future studies could also investigate systems with broken ergodicity, such as jammed or glassy states \cite{charbonneau2012dimensional, charbonneau2014hopping, morse2014geometric}, as well as active matter, which may exhibit non-trivial cage relaxation dynamics \cite{ni2013pushing}. Moreover, exploring systems with soft and anisotropic particles could provide further insights, as these introduce ruggedness into the energy landscapes, even in two-dimensional contexts \cite{yang2016structures, zheng2011glass, mishra2015shape}.

We are grateful for the useful discussions with Sivasurender Chandran, B. Prasanna Venkatesh, Adhip Agarwala, and K. Hima Nagamanasa. We gratefully acknowledge financial support from the Department of Science and Technology (Government of India), INSPIRE fellowship IF200274 (N.H.B.), research initiation grant from IIT Gandhinagar through IP/IITGN/PHY/CM/2021/11 (C.K.M.) and the Start-up Research Grant of Science and Engineering Research Board of Government of India through SRG/2021/001077 (C.K.M.).

\bibliography{references}

\begin{thebibliography}{33}%
\makeatletter
\providecommand \@ifxundefined [1]{%
 \@ifx{#1\undefined}
}%
\providecommand \@ifnum [1]{%
 \ifnum #1\expandafter \@firstoftwo
 \else \expandafter \@secondoftwo
 \fi
}%
\providecommand \@ifx [1]{%
 \ifx #1\expandafter \@firstoftwo
 \else \expandafter \@secondoftwo
 \fi
}%
\providecommand \natexlab [1]{#1}%
\providecommand \enquote  [1]{``#1''}%
\providecommand \bibnamefont  [1]{#1}%
\providecommand \bibfnamefont [1]{#1}%
\providecommand \citenamefont [1]{#1}%
\providecommand \href@noop [0]{\@secondoftwo}%
\providecommand \href [0]{\begingroup \@sanitize@url \@href}%
\providecommand \@href[1]{\@@startlink{#1}\@@href}%
\providecommand \@@href[1]{\endgroup#1\@@endlink}%
\providecommand \@sanitize@url [0]{\catcode `\\12\catcode `\$12\catcode
  `\&12\catcode `\#12\catcode `\^12\catcode `\_12\catcode `\%12\relax}%
\providecommand \@@startlink[1]{}%
\providecommand \@@endlink[0]{}%
\providecommand \url  [0]{\begingroup\@sanitize@url \@url }%
\providecommand \@url [1]{\endgroup\@href {#1}{\urlprefix }}%
\providecommand \urlprefix  [0]{URL }%
\providecommand \Eprint [0]{\href }%
\providecommand \doibase [0]{https://doi.org/}%
\providecommand \selectlanguage [0]{\@gobble}%
\providecommand \bibinfo  [0]{\@secondoftwo}%
\providecommand \bibfield  [0]{\@secondoftwo}%
\providecommand \translation [1]{[#1]}%
\providecommand \BibitemOpen [0]{}%
\providecommand \bibitemStop [0]{}%
\providecommand \bibitemNoStop [0]{.\EOS\space}%
\providecommand \EOS [0]{\spacefactor3000\relax}%
\providecommand \BibitemShut  [1]{\csname bibitem#1\endcsname}%
\let\auto@bib@innerbib\@empty
\bibitem [{\citenamefont {Reis}\ \emph {et~al.}(2007)\citenamefont {Reis},
  \citenamefont {Ingale},\ and\ \citenamefont {Shattuck}}]{reis2007caging}%
  \BibitemOpen
  \bibfield  {author} {\bibinfo {author} {\bibfnamefont {P.~M.}\ \bibnamefont
  {Reis}}, \bibinfo {author} {\bibfnamefont {R.~A.}\ \bibnamefont {Ingale}},\
  and\ \bibinfo {author} {\bibfnamefont {M.~D.}\ \bibnamefont {Shattuck}},\
  }\bibfield  {title} {\bibinfo {title} {Caging dynamics in a granular fluid},\
  }\href@noop {} {\bibfield  {journal} {\bibinfo  {journal} {Phys. Rev. Lett.}\
  }\textbf {\bibinfo {volume} {98}},\ \bibinfo {pages} {188301} (\bibinfo
  {year} {2007})}\BibitemShut {NoStop}%
\bibitem [{\citenamefont {Gallo}\ \emph {et~al.}(1996)\citenamefont {Gallo},
  \citenamefont {Sciortino}, \citenamefont {Tartaglia},\ and\ \citenamefont
  {Chen}}]{gallo1996slow}%
  \BibitemOpen
  \bibfield  {author} {\bibinfo {author} {\bibfnamefont {P.}~\bibnamefont
  {Gallo}}, \bibinfo {author} {\bibfnamefont {F.}~\bibnamefont {Sciortino}},
  \bibinfo {author} {\bibfnamefont {P.}~\bibnamefont {Tartaglia}},\ and\
  \bibinfo {author} {\bibfnamefont {S.-H.}\ \bibnamefont {Chen}},\ }\bibfield
  {title} {\bibinfo {title} {Slow dynamics of water molecules in supercooled
  states},\ }\href@noop {} {\bibfield  {journal} {\bibinfo  {journal} {Phys.
  Rev. Lett.}\ }\textbf {\bibinfo {volume} {76}},\ \bibinfo {pages} {2730}
  (\bibinfo {year} {1996})}\BibitemShut {NoStop}%
\bibitem [{\citenamefont {Sahu}\ \emph {et~al.}(2024)\citenamefont {Sahu},
  \citenamefont {Sharma}, \citenamefont {Schall}, \citenamefont
  {Bhattacharyya},\ and\ \citenamefont {Chikkadi}}]{sahu2024structural}%
  \BibitemOpen
  \bibfield  {author} {\bibinfo {author} {\bibfnamefont {R.}~\bibnamefont
  {Sahu}}, \bibinfo {author} {\bibfnamefont {M.}~\bibnamefont {Sharma}},
  \bibinfo {author} {\bibfnamefont {P.}~\bibnamefont {Schall}}, \bibinfo
  {author} {\bibfnamefont {S.~M.}\ \bibnamefont {Bhattacharyya}},\ and\
  \bibinfo {author} {\bibfnamefont {V.}~\bibnamefont {Chikkadi}},\ }\bibfield
  {title} {\bibinfo {title} {Structural origin of relaxation in dense colloidal
  suspensions},\ }\href@noop {} {\bibfield  {journal} {\bibinfo  {journal}
  {arXiv preprint arXiv:2403.02517}\ } (\bibinfo {year} {2024})}\BibitemShut
  {NoStop}%
\bibitem [{\citenamefont {Li}\ \emph {et~al.}(2020)\citenamefont {Li},
  \citenamefont {Lou}, \citenamefont {Kob},\ and\ \citenamefont
  {Granick}}]{li2020anatomy}%
  \BibitemOpen
  \bibfield  {author} {\bibinfo {author} {\bibfnamefont {B.}~\bibnamefont
  {Li}}, \bibinfo {author} {\bibfnamefont {K.}~\bibnamefont {Lou}}, \bibinfo
  {author} {\bibfnamefont {W.}~\bibnamefont {Kob}},\ and\ \bibinfo {author}
  {\bibfnamefont {S.}~\bibnamefont {Granick}},\ }\bibfield  {title} {\bibinfo
  {title} {Anatomy of cage formation in a two-dimensional glass-forming
  liquid},\ }\href@noop {} {\bibfield  {journal} {\bibinfo  {journal} {Nature}\
  }\textbf {\bibinfo {volume} {587}},\ \bibinfo {pages} {225} (\bibinfo {year}
  {2020})}\BibitemShut {NoStop}%
\bibitem [{\citenamefont {Weeks}\ and\ \citenamefont
  {Weitz}(2002)}]{weeks2002properties}%
  \BibitemOpen
  \bibfield  {author} {\bibinfo {author} {\bibfnamefont {E.~R.}\ \bibnamefont
  {Weeks}}\ and\ \bibinfo {author} {\bibfnamefont {D.}~\bibnamefont {Weitz}},\
  }\bibfield  {title} {\bibinfo {title} {Properties of cage rearrangements
  observed near the colloidal glass transition},\ }\href@noop {} {\bibfield
  {journal} {\bibinfo  {journal} {Phys. Rev. Lett.}\ }\textbf {\bibinfo
  {volume} {89}},\ \bibinfo {pages} {095704} (\bibinfo {year}
  {2002})}\BibitemShut {NoStop}%
\bibitem [{\citenamefont {Rabani}\ \emph {et~al.}(1999)\citenamefont {Rabani},
  \citenamefont {Gezelter},\ and\ \citenamefont {Berne}}]{rabani1999direct}%
  \BibitemOpen
  \bibfield  {author} {\bibinfo {author} {\bibfnamefont {E.}~\bibnamefont
  {Rabani}}, \bibinfo {author} {\bibfnamefont {J.~D.}\ \bibnamefont
  {Gezelter}},\ and\ \bibinfo {author} {\bibfnamefont {B.}~\bibnamefont
  {Berne}},\ }\bibfield  {title} {\bibinfo {title} {Direct observation of
  stretched-exponential relaxation in low-temperature {L}ennard-{J}ones systems
  using the cage correlation function},\ }\href@noop {} {\bibfield  {journal}
  {\bibinfo  {journal} {Phys. Rev. Lett.}\ }\textbf {\bibinfo {volume} {82}},\
  \bibinfo {pages} {3649} (\bibinfo {year} {1999})}\BibitemShut {NoStop}%
\bibitem [{\citenamefont {Quinn}\ and\ \citenamefont
  {Goree}(2002)}]{quinn2002particle}%
  \BibitemOpen
  \bibfield  {author} {\bibinfo {author} {\bibfnamefont {R.}~\bibnamefont
  {Quinn}}\ and\ \bibinfo {author} {\bibfnamefont {J.}~\bibnamefont {Goree}},\
  }\bibfield  {title} {\bibinfo {title} {Particle interaction measurements in a
  {C}oulomb crystal using caged-particle motion},\ }\href@noop {} {\bibfield
  {journal} {\bibinfo  {journal} {Phys. Rev. Lett.}\ }\textbf {\bibinfo
  {volume} {88}},\ \bibinfo {pages} {195001} (\bibinfo {year}
  {2002})}\BibitemShut {NoStop}%
\bibitem [{\citenamefont {Mayer}\ \emph {et~al.}(2008)\citenamefont {Mayer},
  \citenamefont {Zaccarelli}, \citenamefont {Stiakakis}, \citenamefont {Likos},
  \citenamefont {Sciortino}, \citenamefont {Munam}, \citenamefont {Gauthier},
  \citenamefont {Hadjichristidis}, \citenamefont {Iatrou}, \citenamefont
  {Tartaglia} \emph {et~al.}}]{mayer2008asymmetric}%
  \BibitemOpen
  \bibfield  {author} {\bibinfo {author} {\bibfnamefont {C.}~\bibnamefont
  {Mayer}}, \bibinfo {author} {\bibfnamefont {E.}~\bibnamefont {Zaccarelli}},
  \bibinfo {author} {\bibfnamefont {E.}~\bibnamefont {Stiakakis}}, \bibinfo
  {author} {\bibfnamefont {C.}~\bibnamefont {Likos}}, \bibinfo {author}
  {\bibfnamefont {F.}~\bibnamefont {Sciortino}}, \bibinfo {author}
  {\bibfnamefont {A.}~\bibnamefont {Munam}}, \bibinfo {author} {\bibfnamefont
  {M.}~\bibnamefont {Gauthier}}, \bibinfo {author} {\bibfnamefont
  {N.}~\bibnamefont {Hadjichristidis}}, \bibinfo {author} {\bibfnamefont
  {H.}~\bibnamefont {Iatrou}}, \bibinfo {author} {\bibfnamefont
  {P.}~\bibnamefont {Tartaglia}}, \emph {et~al.},\ }\bibfield  {title}
  {\bibinfo {title} {Asymmetric caging in soft colloidal mixtures},\
  }\href@noop {} {\bibfield  {journal} {\bibinfo  {journal} {Nat. Materials}\
  }\textbf {\bibinfo {volume} {7}},\ \bibinfo {pages} {780} (\bibinfo {year}
  {2008})}\BibitemShut {NoStop}%
\bibitem [{\citenamefont {Debets}\ \emph {et~al.}(2021)\citenamefont {Debets},
  \citenamefont {De~Wit},\ and\ \citenamefont {Janssen}}]{debets2021cage}%
  \BibitemOpen
  \bibfield  {author} {\bibinfo {author} {\bibfnamefont {V.~E.}\ \bibnamefont
  {Debets}}, \bibinfo {author} {\bibfnamefont {X.~M.}\ \bibnamefont {De~Wit}},\
  and\ \bibinfo {author} {\bibfnamefont {L.~M.}\ \bibnamefont {Janssen}},\
  }\bibfield  {title} {\bibinfo {title} {Cage length controls the nonmonotonic
  dynamics of active glassy matter},\ }\href@noop {} {\bibfield  {journal}
  {\bibinfo  {journal} {Phys. Rev. Lett.}\ }\textbf {\bibinfo {volume} {127}},\
  \bibinfo {pages} {278002} (\bibinfo {year} {2021})}\BibitemShut {NoStop}%
\bibitem [{\citenamefont {Zhang}\ \emph {et~al.}(2024)\citenamefont {Zhang},
  \citenamefont {Zhang}, \citenamefont {Liu},\ and\ \citenamefont
  {Han}}]{zhang2024anisotropic}%
  \BibitemOpen
  \bibfield  {author} {\bibinfo {author} {\bibfnamefont {H.}~\bibnamefont
  {Zhang}}, \bibinfo {author} {\bibfnamefont {Q.}~\bibnamefont {Zhang}},
  \bibinfo {author} {\bibfnamefont {F.}~\bibnamefont {Liu}},\ and\ \bibinfo
  {author} {\bibfnamefont {Y.}~\bibnamefont {Han}},\ }\bibfield  {title}
  {\bibinfo {title} {Anisotropic-isotropic transition of cages at the glass
  transition},\ }\href@noop {} {\bibfield  {journal} {\bibinfo  {journal}
  {Phys. Rev. Lett.}\ }\textbf {\bibinfo {volume} {132}},\ \bibinfo {pages}
  {078201} (\bibinfo {year} {2024})}\BibitemShut {NoStop}%
\bibitem [{\citenamefont {Illing}\ \emph {et~al.}(2017)\citenamefont {Illing},
  \citenamefont {Fritschi}, \citenamefont {Kaiser}, \citenamefont {Klix},
  \citenamefont {Maret},\ and\ \citenamefont {Keim}}]{illing2017mermin}%
  \BibitemOpen
  \bibfield  {author} {\bibinfo {author} {\bibfnamefont {B.}~\bibnamefont
  {Illing}}, \bibinfo {author} {\bibfnamefont {S.}~\bibnamefont {Fritschi}},
  \bibinfo {author} {\bibfnamefont {H.}~\bibnamefont {Kaiser}}, \bibinfo
  {author} {\bibfnamefont {C.~L.}\ \bibnamefont {Klix}}, \bibinfo {author}
  {\bibfnamefont {G.}~\bibnamefont {Maret}},\ and\ \bibinfo {author}
  {\bibfnamefont {P.}~\bibnamefont {Keim}},\ }\bibfield  {title} {\bibinfo
  {title} {Mermin--wagner fluctuations in {2D} amorphous solids},\ }\href@noop
  {} {\bibfield  {journal} {\bibinfo  {journal} {Proc. Natl. Acad. Sci. USA}\
  }\textbf {\bibinfo {volume} {114}},\ \bibinfo {pages} {1856} (\bibinfo {year}
  {2017})}\BibitemShut {NoStop}%
\bibitem [{\citenamefont {Vivek}\ \emph {et~al.}(2017)\citenamefont {Vivek},
  \citenamefont {Kelleher}, \citenamefont {Chaikin},\ and\ \citenamefont
  {Weeks}}]{vivek2017long}%
  \BibitemOpen
  \bibfield  {author} {\bibinfo {author} {\bibfnamefont {S.}~\bibnamefont
  {Vivek}}, \bibinfo {author} {\bibfnamefont {C.~P.}\ \bibnamefont {Kelleher}},
  \bibinfo {author} {\bibfnamefont {P.~M.}\ \bibnamefont {Chaikin}},\ and\
  \bibinfo {author} {\bibfnamefont {E.~R.}\ \bibnamefont {Weeks}},\ }\bibfield
  {title} {\bibinfo {title} {Long-wavelength fluctuations and the glass
  transition in two dimensions and three dimensions},\ }\href@noop {}
  {\bibfield  {journal} {\bibinfo  {journal} {Proc. Natl. Acad. Sci. USA}\
  }\textbf {\bibinfo {volume} {114}},\ \bibinfo {pages} {1850} (\bibinfo {year}
  {2017})}\BibitemShut {NoStop}%
\bibitem [{\citenamefont {Zheng}\ and\ \citenamefont
  {Earnshaw}(1998)}]{zheng1998lindemann}%
  \BibitemOpen
  \bibfield  {author} {\bibinfo {author} {\bibfnamefont {X.}~\bibnamefont
  {Zheng}}\ and\ \bibinfo {author} {\bibfnamefont {J.}~\bibnamefont
  {Earnshaw}},\ }\bibfield  {title} {\bibinfo {title} {On the {L}indemann
  criterion in {2D}},\ }\href@noop {} {\bibfield  {journal} {\bibinfo
  {journal} {Europhys. Lett.}\ }\textbf {\bibinfo {volume} {41}},\ \bibinfo
  {pages} {635} (\bibinfo {year} {1998})}\BibitemShut {NoStop}%
\bibitem [{\citenamefont {Shiba}\ \emph {et~al.}(2018)\citenamefont {Shiba},
  \citenamefont {Keim},\ and\ \citenamefont {Kawasaki}}]{Shiba_2018}%
  \BibitemOpen
  \bibfield  {author} {\bibinfo {author} {\bibfnamefont {H.}~\bibnamefont
  {Shiba}}, \bibinfo {author} {\bibfnamefont {P.}~\bibnamefont {Keim}},\ and\
  \bibinfo {author} {\bibfnamefont {T.}~\bibnamefont {Kawasaki}},\ }\bibfield
  {title} {\bibinfo {title} {Isolating long-wavelength fluctuation from
  structural relaxation in two-dimensional glass: cage-relative displacement},\
  }\href@noop {} {\bibfield  {journal} {\bibinfo  {journal} {J. Phys.: Condens.
  Matter}\ }\textbf {\bibinfo {volume} {30}},\ \bibinfo {pages} {094004}
  (\bibinfo {year} {2018})}\BibitemShut {NoStop}%
\bibitem [{\citenamefont {Flenner}\ and\ \citenamefont
  {Szamel}(2015)}]{flenner2015fundamental}%
  \BibitemOpen
  \bibfield  {author} {\bibinfo {author} {\bibfnamefont {E.}~\bibnamefont
  {Flenner}}\ and\ \bibinfo {author} {\bibfnamefont {G.}~\bibnamefont
  {Szamel}},\ }\bibfield  {title} {\bibinfo {title} {Fundamental differences
  between glassy dynamics in two and three dimensions},\ }\href@noop {}
  {\bibfield  {journal} {\bibinfo  {journal} {Nature communications}\ }\textbf
  {\bibinfo {volume} {6}},\ \bibinfo {pages} {7392} (\bibinfo {year}
  {2015})}\BibitemShut {NoStop}%
\bibitem [{\citenamefont {Li}\ \emph {et~al.}(2019)\citenamefont {Li},
  \citenamefont {Mishra}, \citenamefont {Sun}, \citenamefont {Zhao},
  \citenamefont {Mason}, \citenamefont {Ganapathy},\ and\ \citenamefont
  {Pica~Ciamarra}}]{li2019long}%
  \BibitemOpen
  \bibfield  {author} {\bibinfo {author} {\bibfnamefont {Y.-W.}\ \bibnamefont
  {Li}}, \bibinfo {author} {\bibfnamefont {C.~K.}\ \bibnamefont {Mishra}},
  \bibinfo {author} {\bibfnamefont {Z.-Y.}\ \bibnamefont {Sun}}, \bibinfo
  {author} {\bibfnamefont {K.}~\bibnamefont {Zhao}}, \bibinfo {author}
  {\bibfnamefont {T.~G.}\ \bibnamefont {Mason}}, \bibinfo {author}
  {\bibfnamefont {R.}~\bibnamefont {Ganapathy}},\ and\ \bibinfo {author}
  {\bibfnamefont {M.}~\bibnamefont {Pica~Ciamarra}},\ }\bibfield  {title}
  {\bibinfo {title} {Long-wavelength fluctuations and anomalous dynamics in
  2-dimensional liquids},\ }\href@noop {} {\bibfield  {journal} {\bibinfo
  {journal} {Proc. Natl. Acad. Sci. USA}\ }\textbf {\bibinfo {volume} {116}},\
  \bibinfo {pages} {22977} (\bibinfo {year} {2019})}\BibitemShut {NoStop}%
\bibitem [{\citenamefont {Shiba}\ \emph {et~al.}(2019)\citenamefont {Shiba},
  \citenamefont {Kawasaki},\ and\ \citenamefont {Kim}}]{shiba2019local}%
  \BibitemOpen
  \bibfield  {author} {\bibinfo {author} {\bibfnamefont {H.}~\bibnamefont
  {Shiba}}, \bibinfo {author} {\bibfnamefont {T.}~\bibnamefont {Kawasaki}},\
  and\ \bibinfo {author} {\bibfnamefont {K.}~\bibnamefont {Kim}},\ }\bibfield
  {title} {\bibinfo {title} {Local density fluctuation governs the divergence
  of viscosity underlying elastic and hydrodynamic anomalies in a 2d
  glass-forming liquid},\ }\href@noop {} {\bibfield  {journal} {\bibinfo
  {journal} {Phys. Rev. Lett.}\ }\textbf {\bibinfo {volume} {123}},\ \bibinfo
  {pages} {265501} (\bibinfo {year} {2019})}\BibitemShut {NoStop}%
\bibitem [{\citenamefont {Barbhuiya}\ \emph {et~al.}(2023)\citenamefont
  {Barbhuiya}, \citenamefont {Yodh},\ and\ \citenamefont
  {Mishra}}]{barbhuiya2023direction}%
  \BibitemOpen
  \bibfield  {author} {\bibinfo {author} {\bibfnamefont {N.~H.}\ \bibnamefont
  {Barbhuiya}}, \bibinfo {author} {\bibfnamefont {A.}~\bibnamefont {Yodh}},\
  and\ \bibinfo {author} {\bibfnamefont {C.~K.}\ \bibnamefont {Mishra}},\
  }\bibfield  {title} {\bibinfo {title} {Direction-dependent dynamics of
  colloidal particle pairs and the {S}tokes-{E}instein relation in
  quasi-two-dimensional fluids},\ }\href@noop {} {\bibfield  {journal}
  {\bibinfo  {journal} {Nat. Communications}\ }\textbf {\bibinfo {volume}
  {14}},\ \bibinfo {pages} {5109} (\bibinfo {year} {2023})}\BibitemShut
  {NoStop}%
\bibitem [{\citenamefont {Crocker}\ and\ \citenamefont
  {Grier}(1996)}]{crocker1996methods}%
  \BibitemOpen
  \bibfield  {author} {\bibinfo {author} {\bibfnamefont {J.~C.}\ \bibnamefont
  {Crocker}}\ and\ \bibinfo {author} {\bibfnamefont {D.~G.}\ \bibnamefont
  {Grier}},\ }\bibfield  {title} {\bibinfo {title} {Methods of digital video
  microscopy for colloidal studies},\ }\href@noop {} {\bibfield  {journal}
  {\bibinfo  {journal} {J. Colloid Interface Sci.}\ }\textbf {\bibinfo {volume}
  {179}},\ \bibinfo {pages} {298} (\bibinfo {year} {1996})}\BibitemShut
  {NoStop}%
\bibitem [{\citenamefont {Burrough}\ \emph {et~al.}(2015)\citenamefont
  {Burrough}, \citenamefont {McDonnell},\ and\ \citenamefont
  {Lloyd}}]{burrough2015principles}%
  \BibitemOpen
  \bibfield  {author} {\bibinfo {author} {\bibfnamefont {P.~A.}\ \bibnamefont
  {Burrough}}, \bibinfo {author} {\bibfnamefont {R.~A.}\ \bibnamefont
  {McDonnell}},\ and\ \bibinfo {author} {\bibfnamefont {C.~D.}\ \bibnamefont
  {Lloyd}},\ }\href@noop {} {\emph {\bibinfo {title} {Principles of
  geographical information systems}}}\ (\bibinfo  {publisher} {Oxford
  University Press, USA},\ \bibinfo {year} {2015})\BibitemShut {NoStop}%
\bibitem [{sup()}]{supplemental}%
  \BibitemOpen
  \href@noop {} {}\bibinfo {note} {Method for determining the inner contour of
  the probability distribution's maxima through ray tracing: Rays are emanated
  in all directions and traced from the probability distribution's center
  (mean) to identify points where the probability is maximum along the ray
  directions. These points define the contour on which an ellipse is
  fitted.}\BibitemShut {Stop}%
\bibitem [{\citenamefont {Chen}\ \emph {et~al.}(2023)\citenamefont {Chen},
  \citenamefont {Ye}, \citenamefont {Wang}, \citenamefont {Huang},
  \citenamefont {Tong}, \citenamefont {Jin}, \citenamefont {Chen},
  \citenamefont {Tanaka},\ and\ \citenamefont {Tan}}]{chen2023visualizing}%
  \BibitemOpen
  \bibfield  {author} {\bibinfo {author} {\bibfnamefont {Y.}~\bibnamefont
  {Chen}}, \bibinfo {author} {\bibfnamefont {Z.}~\bibnamefont {Ye}}, \bibinfo
  {author} {\bibfnamefont {K.}~\bibnamefont {Wang}}, \bibinfo {author}
  {\bibfnamefont {J.}~\bibnamefont {Huang}}, \bibinfo {author} {\bibfnamefont
  {H.}~\bibnamefont {Tong}}, \bibinfo {author} {\bibfnamefont {Y.}~\bibnamefont
  {Jin}}, \bibinfo {author} {\bibfnamefont {K.}~\bibnamefont {Chen}}, \bibinfo
  {author} {\bibfnamefont {H.}~\bibnamefont {Tanaka}},\ and\ \bibinfo {author}
  {\bibfnamefont {P.}~\bibnamefont {Tan}},\ }\bibfield  {title} {\bibinfo
  {title} {Visualizing slow internal relaxations in a two-dimensional glassy
  system},\ }\href@noop {} {\bibfield  {journal} {\bibinfo  {journal} {Nat.
  Phys.}\ }\textbf {\bibinfo {volume} {19}},\ \bibinfo {pages} {969} (\bibinfo
  {year} {2023})}\BibitemShut {NoStop}%
\bibitem [{\citenamefont {Cao}\ \emph {et~al.}(2017)\citenamefont {Cao},
  \citenamefont {Zhang},\ and\ \citenamefont {Han}}]{cao2017release}%
  \BibitemOpen
  \bibfield  {author} {\bibinfo {author} {\bibfnamefont {X.}~\bibnamefont
  {Cao}}, \bibinfo {author} {\bibfnamefont {H.}~\bibnamefont {Zhang}},\ and\
  \bibinfo {author} {\bibfnamefont {Y.}~\bibnamefont {Han}},\ }\bibfield
  {title} {\bibinfo {title} {Release of free-volume bubbles by
  cooperative-rearrangement regions during the deposition growth of a colloidal
  glass},\ }\href@noop {} {\bibfield  {journal} {\bibinfo  {journal} {Nat.
  Communications}\ }\textbf {\bibinfo {volume} {8}},\ \bibinfo {pages} {362}
  (\bibinfo {year} {2017})}\BibitemShut {NoStop}%
\bibitem [{\citenamefont {van Baalen}\ \emph {et~al.}(2023)\citenamefont {van
  Baalen}, \citenamefont {Vialetto},\ and\ \citenamefont
  {Isa}}]{van2023tuning}%
  \BibitemOpen
  \bibfield  {author} {\bibinfo {author} {\bibfnamefont {C.}~\bibnamefont {van
  Baalen}}, \bibinfo {author} {\bibfnamefont {J.}~\bibnamefont {Vialetto}},\
  and\ \bibinfo {author} {\bibfnamefont {L.}~\bibnamefont {Isa}},\ }\bibfield
  {title} {\bibinfo {title} {Tuning electrostatic interactions of colloidal
  particles at oil-water interfaces with organic salts},\ }\href@noop {}
  {\bibfield  {journal} {\bibinfo  {journal} {Phys. Rev. Lett.}\ }\textbf
  {\bibinfo {volume} {131}},\ \bibinfo {pages} {128202} (\bibinfo {year}
  {2023})}\BibitemShut {NoStop}%
\bibitem [{\citenamefont {Bausch}\ \emph {et~al.}(2003)\citenamefont {Bausch},
  \citenamefont {Bowick}, \citenamefont {Cacciuto}, \citenamefont {Dinsmore},
  \citenamefont {Hsu}, \citenamefont {Nelson}, \citenamefont {Nikolaides},
  \citenamefont {Travesset},\ and\ \citenamefont {Weitz}}]{bausch2003grain}%
  \BibitemOpen
  \bibfield  {author} {\bibinfo {author} {\bibfnamefont {A.}~\bibnamefont
  {Bausch}}, \bibinfo {author} {\bibfnamefont {M.~J.}\ \bibnamefont {Bowick}},
  \bibinfo {author} {\bibfnamefont {A.}~\bibnamefont {Cacciuto}}, \bibinfo
  {author} {\bibfnamefont {A.}~\bibnamefont {Dinsmore}}, \bibinfo {author}
  {\bibfnamefont {M.}~\bibnamefont {Hsu}}, \bibinfo {author} {\bibfnamefont
  {D.}~\bibnamefont {Nelson}}, \bibinfo {author} {\bibfnamefont
  {M.}~\bibnamefont {Nikolaides}}, \bibinfo {author} {\bibfnamefont
  {A.}~\bibnamefont {Travesset}},\ and\ \bibinfo {author} {\bibfnamefont
  {D.}~\bibnamefont {Weitz}},\ }\bibfield  {title} {\bibinfo {title} {Grain
  boundary scars and spherical crystallography},\ }\href@noop {} {\bibfield
  {journal} {\bibinfo  {journal} {Science}\ }\textbf {\bibinfo {volume}
  {299}},\ \bibinfo {pages} {1716} (\bibinfo {year} {2003})}\BibitemShut
  {NoStop}%
\bibitem [{\citenamefont {Singh}\ \emph {et~al.}(2022)\citenamefont {Singh},
  \citenamefont {Sood},\ and\ \citenamefont
  {Ganapathy}}]{singh2022observation}%
  \BibitemOpen
  \bibfield  {author} {\bibinfo {author} {\bibfnamefont {N.}~\bibnamefont
  {Singh}}, \bibinfo {author} {\bibfnamefont {A.}~\bibnamefont {Sood}},\ and\
  \bibinfo {author} {\bibfnamefont {R.}~\bibnamefont {Ganapathy}},\ }\bibfield
  {title} {\bibinfo {title} {Observation of two-step melting on a sphere},\
  }\href@noop {} {\bibfield  {journal} {\bibinfo  {journal} {Proc. Natl. Acad.
  Sci. USA}\ }\textbf {\bibinfo {volume} {119}},\ \bibinfo {pages}
  {e2206470119} (\bibinfo {year} {2022})}\BibitemShut {NoStop}%
\bibitem [{\citenamefont {Charbonneau}\ \emph {et~al.}(2012)\citenamefont
  {Charbonneau}, \citenamefont {Ikeda}, \citenamefont {Parisi},\ and\
  \citenamefont {Zamponi}}]{charbonneau2012dimensional}%
  \BibitemOpen
  \bibfield  {author} {\bibinfo {author} {\bibfnamefont {P.}~\bibnamefont
  {Charbonneau}}, \bibinfo {author} {\bibfnamefont {A.}~\bibnamefont {Ikeda}},
  \bibinfo {author} {\bibfnamefont {G.}~\bibnamefont {Parisi}},\ and\ \bibinfo
  {author} {\bibfnamefont {F.}~\bibnamefont {Zamponi}},\ }\bibfield  {title}
  {\bibinfo {title} {Dimensional study of the caging order parameter at the
  glass transition},\ }\href@noop {} {\bibfield  {journal} {\bibinfo  {journal}
  {Proc. Natl. Acad. Sci. USA}\ }\textbf {\bibinfo {volume} {109}},\ \bibinfo
  {pages} {13939} (\bibinfo {year} {2012})}\BibitemShut {NoStop}%
\bibitem [{\citenamefont {Charbonneau}\ \emph {et~al.}(2014)\citenamefont
  {Charbonneau}, \citenamefont {Jin}, \citenamefont {Parisi},\ and\
  \citenamefont {Zamponi}}]{charbonneau2014hopping}%
  \BibitemOpen
  \bibfield  {author} {\bibinfo {author} {\bibfnamefont {P.}~\bibnamefont
  {Charbonneau}}, \bibinfo {author} {\bibfnamefont {Y.}~\bibnamefont {Jin}},
  \bibinfo {author} {\bibfnamefont {G.}~\bibnamefont {Parisi}},\ and\ \bibinfo
  {author} {\bibfnamefont {F.}~\bibnamefont {Zamponi}},\ }\bibfield  {title}
  {\bibinfo {title} {Hopping and the {S}tokes-{E}instein relation breakdown in
  simple glass formers},\ }\href@noop {} {\bibfield  {journal} {\bibinfo
  {journal} {Proc. Natl. Acad. Sci. USA}\ }\textbf {\bibinfo {volume} {111}},\
  \bibinfo {pages} {15025} (\bibinfo {year} {2014})}\BibitemShut {NoStop}%
\bibitem [{\citenamefont {Morse}\ and\ \citenamefont
  {Corwin}(2014)}]{morse2014geometric}%
  \BibitemOpen
  \bibfield  {author} {\bibinfo {author} {\bibfnamefont {P.~K.}\ \bibnamefont
  {Morse}}\ and\ \bibinfo {author} {\bibfnamefont {E.~I.}\ \bibnamefont
  {Corwin}},\ }\bibfield  {title} {\bibinfo {title} {Geometric signatures of
  jamming in the mechanical vacuum},\ }\href@noop {} {\bibfield  {journal}
  {\bibinfo  {journal} {Phys. Rev. Lett.}\ }\textbf {\bibinfo {volume} {112}},\
  \bibinfo {pages} {115701} (\bibinfo {year} {2014})}\BibitemShut {NoStop}%
\bibitem [{\citenamefont {Ni}\ \emph {et~al.}(2013)\citenamefont {Ni},
  \citenamefont {Stuart},\ and\ \citenamefont {Dijkstra}}]{ni2013pushing}%
  \BibitemOpen
  \bibfield  {author} {\bibinfo {author} {\bibfnamefont {R.}~\bibnamefont
  {Ni}}, \bibinfo {author} {\bibfnamefont {M.~A.~C.}\ \bibnamefont {Stuart}},\
  and\ \bibinfo {author} {\bibfnamefont {M.}~\bibnamefont {Dijkstra}},\
  }\bibfield  {title} {\bibinfo {title} {Pushing the glass transition towards
  random close packing using self-propelled hard spheres},\ }\href@noop {}
  {\bibfield  {journal} {\bibinfo  {journal} {Nat. Communications}\ }\textbf
  {\bibinfo {volume} {4}},\ \bibinfo {pages} {2704} (\bibinfo {year}
  {2013})}\BibitemShut {NoStop}%
\bibitem [{\citenamefont {Yang}\ \emph {et~al.}(2016)\citenamefont {Yang},
  \citenamefont {Liu}, \citenamefont {Yang}, \citenamefont {Wang},\ and\
  \citenamefont {Chen}}]{yang2016structures}%
  \BibitemOpen
  \bibfield  {author} {\bibinfo {author} {\bibfnamefont {X.}~\bibnamefont
  {Yang}}, \bibinfo {author} {\bibfnamefont {R.}~\bibnamefont {Liu}}, \bibinfo
  {author} {\bibfnamefont {M.}~\bibnamefont {Yang}}, \bibinfo {author}
  {\bibfnamefont {W.-H.}\ \bibnamefont {Wang}},\ and\ \bibinfo {author}
  {\bibfnamefont {K.}~\bibnamefont {Chen}},\ }\bibfield  {title} {\bibinfo
  {title} {Structures of local rearrangements in soft colloidal glasses},\
  }\href@noop {} {\bibfield  {journal} {\bibinfo  {journal} {Phys. Rev. Lett.}\
  }\textbf {\bibinfo {volume} {116}},\ \bibinfo {pages} {238003} (\bibinfo
  {year} {2016})}\BibitemShut {NoStop}%
\bibitem [{\citenamefont {Zheng}\ \emph {et~al.}(2011)\citenamefont {Zheng},
  \citenamefont {Wang},\ and\ \citenamefont {Han}}]{zheng2011glass}%
  \BibitemOpen
  \bibfield  {author} {\bibinfo {author} {\bibfnamefont {Z.}~\bibnamefont
  {Zheng}}, \bibinfo {author} {\bibfnamefont {F.}~\bibnamefont {Wang}},\ and\
  \bibinfo {author} {\bibfnamefont {Y.}~\bibnamefont {Han}},\ }\bibfield
  {title} {\bibinfo {title} {Glass transitions in quasi-two-dimensional
  suspensions of colloidal ellipsoids},\ }\href@noop {} {\bibfield  {journal}
  {\bibinfo  {journal} {Phys. Rev. Lett.}\ }\textbf {\bibinfo {volume} {107}},\
  \bibinfo {pages} {065702} (\bibinfo {year} {2011})}\BibitemShut {NoStop}%
\bibitem [{\citenamefont {Mishra}\ and\ \citenamefont
  {Ganapathy}(2015)}]{mishra2015shape}%
  \BibitemOpen
  \bibfield  {author} {\bibinfo {author} {\bibfnamefont {C.~K.}\ \bibnamefont
  {Mishra}}\ and\ \bibinfo {author} {\bibfnamefont {R.}~\bibnamefont
  {Ganapathy}},\ }\bibfield  {title} {\bibinfo {title} {Shape of dynamical
  heterogeneities and fractional {S}tokes-{E}instein and
  {S}tokes-{E}instein-{D}ebye relations in quasi-two-dimensional suspensions of
  colloidal ellipsoids},\ }\href@noop {} {\bibfield  {journal} {\bibinfo
  {journal} {Phys. Rev. Lett.}\ }\textbf {\bibinfo {volume} {114}},\ \bibinfo
  {pages} {198302} (\bibinfo {year} {2015})}\BibitemShut {NoStop}%
\end{thebibliography}%
\end{document}